\documentclass{an}

\usepackage{times}
\usepackage{graphicx}

\begin{document}
\sloppy

\title{Delta Scuti stars in the Praesepe cluster observed by the MOST\thanks{Based on data from the {\it MOST} satellite, a Canadian Space Agency mission, jointly operated by Dynacon Inc., the University of Toronto Institute for Aerospace Studies and the University of British Columbia with the assistance of the University of Vienna.} satellite}

\author{M. Breger\inst{1,2}$^{,}$\thanks{Corresponding author: \email{michel.breger@univie.ac.at}} \and
M. Hareter\inst{1} \and
Markus Endl\inst{1} \and
R. Kuschnig\inst{1} \and
W. W. Weiss\inst{1} \and
J. M. Matthews\inst{3} \and
D. B. Guenther\inst{4}\and 
A. F. J. Moffat\inst{5} \and
J. F. Rowe\inst{6} \and
S. M. Rucinski\inst{7} \and
D. Sasselov\inst{8}}

\institute{
Institut f\"ur Astronomie, Universit\"at Wien,
T\"urkenschanzstrasse 17, A-1180 Vienna, Austria \\
\email michel.breger@univie.ac.at  \and
Department of Astronomy, University of Texas, Austin, TX 78712, USA  \and
Department of Physics and Astronomy, University of British Columbia, 6224 Agricultural Road, Vancouver, BC V6T 1Z1, Canada \and 
Department of Astronomy and Physics, St. Mary's University, Halifax, NS B3H 3C3, Canada \and
D\'epartment de physique, Universit\'e de Montr\'eal, C.P.6128, Succ. Centre-Ville, Montr\'eal, QC H3C 3J7, Canada \and
NASA-Ames Research Park, MS-244-30, Moffett Field, CA 94035, USA \and 
Department of Astronomy \& Astrophysics, University of Toronto, 50 St. George Street, Toronto, ON  M5S 3H4 \and  
Harvard-Smithsonian Center for Astrophysics, 60 Garden Street, Cambridge, MA 02138, USA
}


\keywords{delta Scuti stars -- open clusters and associations: individual (Praesepe) -- stars: individual (BS Cnc, BT Cnc, EP Cnc, HD 73872) -- stars: oscillations --  techniques: photometric}

\abstract
{The Praesepe cluster contains a number of $\delta$~Sct and $\gamma$~Dor pulsators.
Asteroseismology of cluster stars is simplified by the common distance, age and stellar abundances. Since asteroseismology requires a
large number of known frequencies, the small pulsation amplitudes of these stars require space satellite campaigns.
The present study utilizes photometric {{\tt MOST}} satellite measurements in order to determine the pulsation frequencies
of two evolved (EP Cnc, BT Cnc) and two main-sequence (BS Cnc, HD 73872) $\delta$~Sct stars in the Praesepe cluster.
The frequency analysis of the 2008 and 2009 data detected up to 34 frequencies per star with most amplitudes in the submillimag range. In BS Cnc, two modes showed strong amplitude variability between 2008 and 2009. The frequencies ranged from 0.76 to 41.7 cd$^{-1}$. After considering the different evolutionary states and mean stellar densities of these four stars, the differences and large ranges in frequency remain.}

\maketitle
\titlerunning{Praesepe Delta Scuti stars}
\authorrunning{M. Breger et al.}

\section{Introduction}
The majority of stars of spectral type A and F pulsate with pressure (p) and/or gravity (g) modes. The stars on and near the main sequence with
dominant pressure modes are known as $\delta$~Sct stars, while the (usually cooler) stars with dominant gravity modes are known as $\gamma$ Dor stars.
Typical periods of $\delta$~Sct stars range from 20 minutes to 8 hours for the dominant modes. Recent campaigns have revealed hundreds of simultaneously excited pulsation modes in some of these stars. The vast majority of the pulsations are low-amplitude nonradial modes.
The new space-satellite {\tt Kepler} measurements have clearly confirmed the previous discovery (e.g., Handler 2009) that many of the A/F star pulsators near the main sequence are hybrid pulsators showing both pressure and longer-period gravity modes. 

Besides the 'standard' low-amplitude multimode pulsators, which pulsate with hundreds of nonradial pulsation modes, a $\delta$~Sct subgroup is usually recognized: the HADS (high-amplitude Delta Scuti) stars are slow rotators with one or more dominant radial modes. The commonly used definition of HADS stars in terms of peak-to-peak amplitudes in excess of 0.3 mag is not meaningful in the case of multimode pulsators. Also, recent measurements have shown that slowly rotating stars with one or two dominant radial modes, which are the characteristic of HADS stars, may even have smaller amplitudes than this artificial limit, e.g., the {\tt Kepler} star KIC 9700322 (Breger et al. 2011).
The sizes of the amplitudes and the determination of whether the dominant modes are radial or nonradial are a strong function of the rotational velocity without an absolute limiting value (Breger 2000). Consequently, it is difficult to define (rather than just describe) the HADS subgroup.

Several variability surveys in the Praesepe cluster have been undertaken (e.g., Breger 1970; Frandsen et al. 2001, Dall et al. 2002) and have revealed a number of
low-ampli\-tude $\delta$~Sct stars. The pulsations of these stars
are of interest, because as cluster members they share similar properties such as chemical abundance and their distances are accurately known. This simplifies
the asteroseismic modeling. In this paper we examine the pulsation properties of four pulsators studied with the {\tt MOST} satellite. Here we note that {\tt MOST} satellite measurements
were already successfully used in the study of the spectroscopic binary HD 73709 by Pribulla et al. (2009, 2010).

In this paper we present the new results for four pulsating members of the Praesepe cluster. More detailed references to previous investigations of these stars are given at the end of the discussion of each star together with a comparison of the results.

\section{MOST observations and reductions}

\begin{figure*}[htb]
\centering
\includegraphics[width=\textwidth]{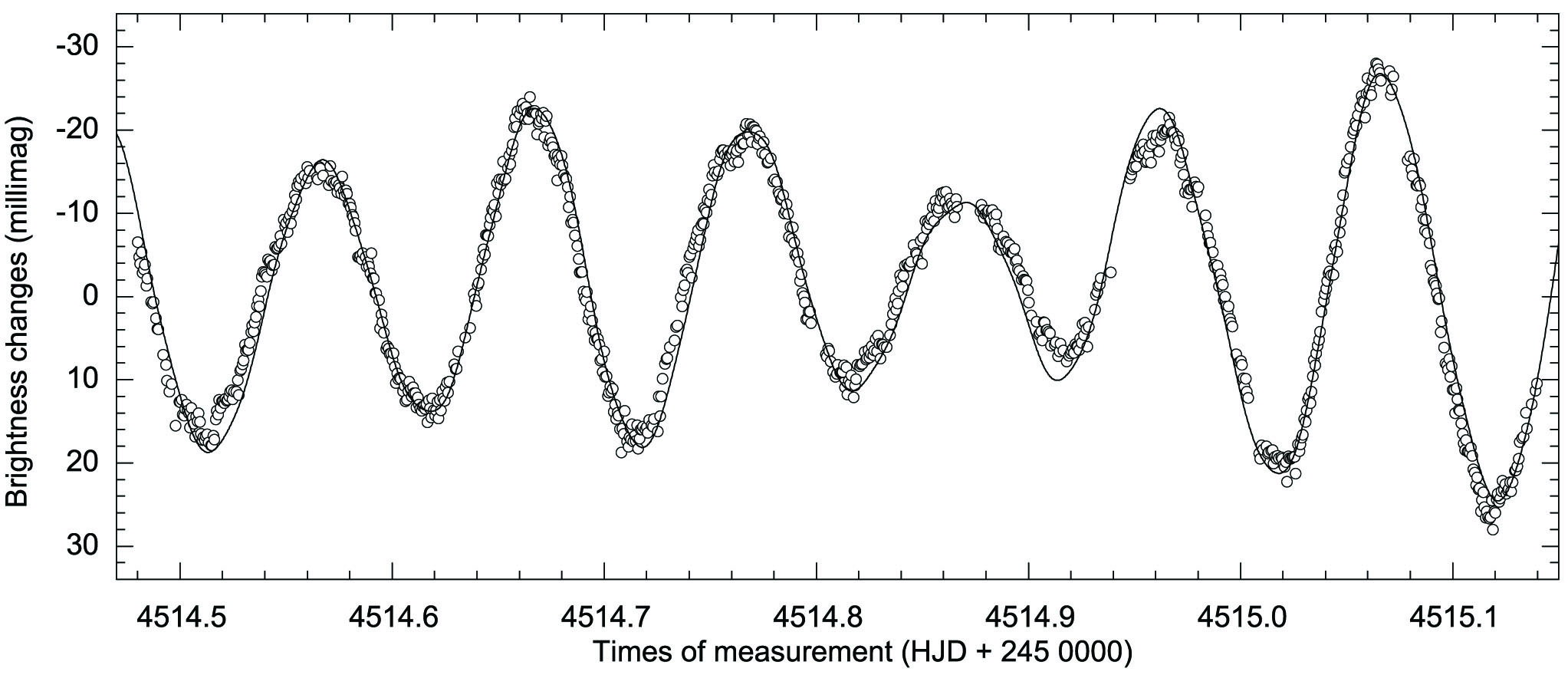} 
\caption{Sample light curve of BT Cnc covering two-thirds of a day. The continuous curve is the fit derived in this paper.}
\label{btic}
\end{figure*}

MOST (Microvariability and Oscillations of STars,  Walker et al. 2003) is a small Canadian spacecraft with a 15-cm telescope designed for photometry of bright stars. Due to the polar dusk-dawn orbit, continuous photometry for about one month can be obtained for stars close to the ecliptic. The nearby open cluster Praesepe was observed from 2008 January 24 to February 25, and from 2009 February 12 to March 1. MOST employs a custom broadband filter with a wavelength range 350 to 750\,nm.

The Direct Imaging (DI) and guide star data were used for this investigation. The data were reduced using the data reduction developed by Huber \& Reegen (2008) and by Hareter (2008).
The basic function of these data reduction programs is based on the decorrelation technique for MOST main targets (Reegen et al. 2006), modified to meet the requirements of the particular data formats. The DI data are available as 20x20 pixel images centered on the particular star. The guide star data 
are a compressed version of this format, i.e., the intensities of the pixels are added, where the mean background is subtracted on-board. Each measurement consists of this sum and the corresponding integration time only. (In contrast to the DI data, no images of the guide stars are available). In spite of this on-board background subtraction, the data still contain heavy stray light which requires additional data reduction. 

The customary three-star photometry reduction scheme cannot be applied, since the stray light is distributed very inhomogeneously on the CCD chip. The mean background value, which is subtracted on-board, is not stored due to downlink limitations; hence the true raw data cannot be reconstructed. To improve the accuracy of the data, it is possible to use the decorrelation technique, as done by the previously mentioned data reduction techniques. Here constant comparison stars are used instead of background pixels, whose individual values are not available. 
    
During 2008, four $\delta$~Sct stars (BT Cnc, EP Cnc, BS Cnc and HD 73872) could be observed. The measurements, consisting of about 34250 individual measurements for each star, covered 32d. During 2009, EP Cnc and BS Cnc were measured again with about 17000 measurements per star, covering 17d.

\section{Frequency analyses}

The frequency analysis was conducted using the  {\tt 
PERIOD04} software package (Lenz \& Breger 2005)
 which utilizes single-frequency Fourier as well
as multifrequency least-squares algorithms. The latter technique fits a number of
simultaneous sinusoidal variations in the magnitude domain and does not rely on prewhitening. The multifrequency solution obtained in this manner needs to be examined in detail in order to (i) determine the statistical significance of each detected frequency, and (ii) separate the frequencies intrinsic to the star from the instrumental effects caused by the MOST satellite.

The statistical significance of the detected frequencies was determined from the amplitude signal/noise (S/N) ratio, which was calculated in the following manner:
for each possible new frequency, $f(test)$, the amplitude of $f(test)$ was derived from a multifrequency solution with $f(test)$ included as one of the frequencies. The noise was determined from the best multifrequency solution using all the detected significant frequencies $excluding$ $f(test)$. This solution was then prewhitened and the average noise calculated from the amplitude spectrum using a 5~cd$^{-1}$ range centered around the frequency under consideration. This made it possible to derive an amplitude signal/noise ratio without biassing the outcome by assuming the significance of $f(test)$. Furthermore, where data were available from both 2008 and 2009, the signal/noise ratio was determined from a common solution. This guards against any hypothetical spurious peaks occurring in only a single data set. Following Breger et al. (1993)
 and Kuschnig et al. (1997), a frequency was regarded as significant if the amplitude signal/noise ratio was 4.0 or greater.
The final residuals between the data and the fits were between 2.3 and 3.1 millimag per single measurement.

We regard our chosen method to include or reject frequency peaks, based on their amplitude signal/noise ratio, as conservative. The chief remaining uncertainty concerns any unrecognized systematic instrumental (orbital) effects.

\begin{figure}[htb]
\centering
\includegraphics[width=\columnwidth]{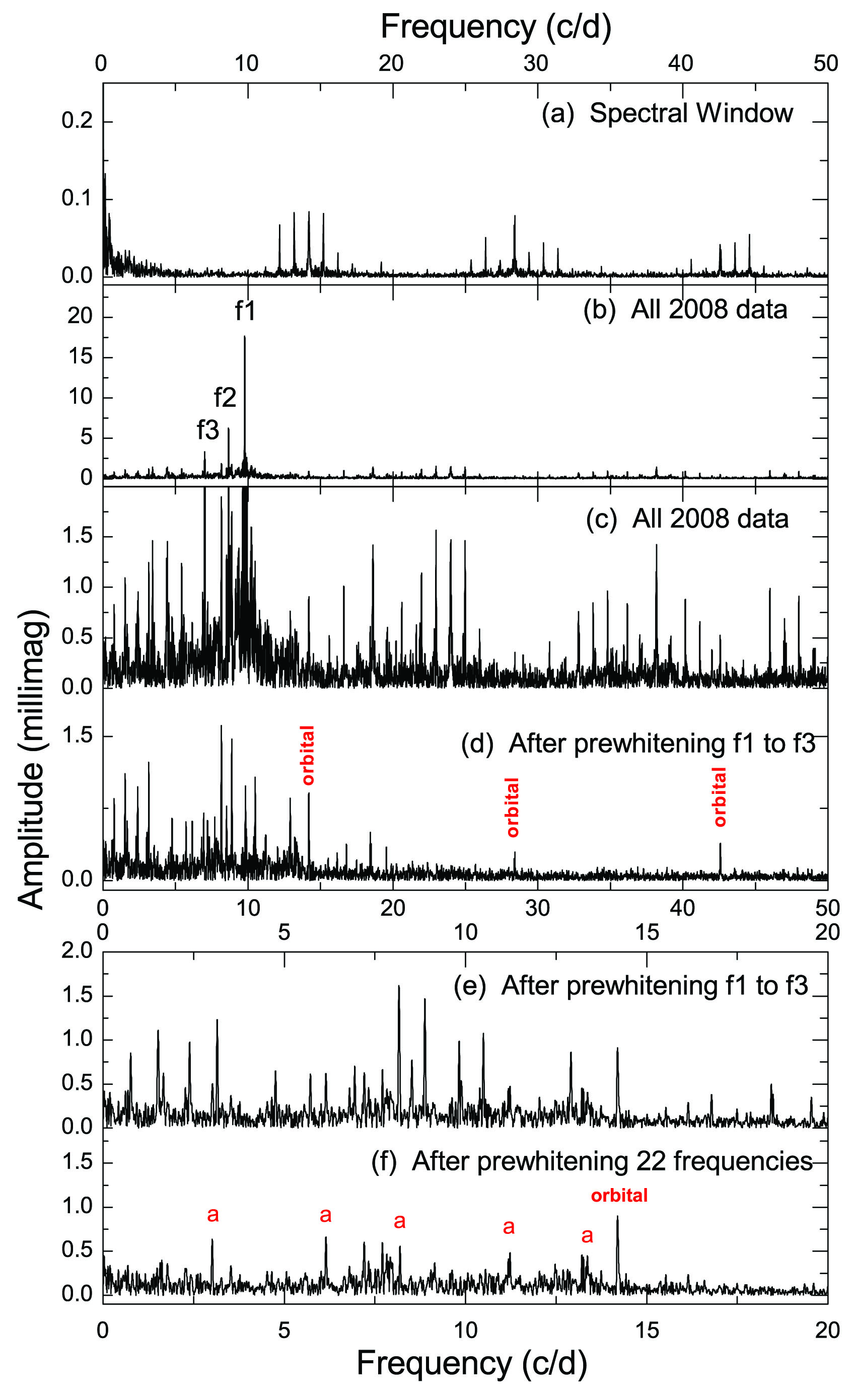}
\caption{Amplitude spectra of BT Cnc. The spectral window is presented at the top. Panel (c) shows the low-amplitude peaks of the 2008 data (with ``f1'' to ``f3'' off scale).
Panel (d) shows that most peaks in panel (c) disappear after prewhitening the three dominant frequencies. At frequencies above 20 cd$^{-1}$, only the orbital frequencies of the satellite are left. Panels (e) and (f) demonstrate that after prewhitening the 22 detected stellar frequencies, most of the remaining peaks are instrumental artifacts (marked by 'a').}
\label{btcnc}
\end{figure}

The remaining stray-light artifacts are far from sinusoidal and not of constant amplitude in each cycle. Hence, the stray light frequencies cannot simply be prewhitened. A further difficulty arises from the outlier rejection, which are most abundant at the phases where the stray light is strongest. Moreover, there is also a modulation of the stray light with one cycle per day. This leads to a rather complicated aliasing pattern (spectral window), where the aliasing frequencies are f$_\mathrm{orb}$, 2f$_\mathrm{orb}$, surrounded by aliasing with the modulation frequencies, which is shown in Fig. \ref{btcnc} top panel. These pattern also appears in the Fourier transform of a MOST data set, not only in the spectral window.
Further known artifacts of the MOST photometry can be found at 1, 2, 3 cd$^{-1}$, usually with decreasing amplitudes.

\section{BT Cnc (= 38 Cnc = HD 73575 = KW 204)}

The data of the star BT Cnc are typical of all four stars. Consequently, we can use the 2008 results for BT Cnc to illustrate our analysis and techniques.
Fig. \ref{btic} shows a small part of the light curve together with the fit derived below and is typical of BT Cnc and the other stars. The small gaps in the coverage and the small systematic deviations between the fit and the observations are caused by instrumental problems of the satellite.
\begin{table}[htb]
\caption{Pulsation frequencies of BT Cnc for 2008}
\label{freqs}
\begin{center}
\begin{tabular}{lrcrr}
\hline
\noalign{\smallskip}
\multicolumn{3}{c}{Frequency} & Amplitude & S/N\\
& cd$^{-1}$ & & millimag & ratio\\
\noalign{\smallskip}
\hline
\noalign{\smallskip}
& $\pm$ 0.001 & & $\pm$ 0.03\\
\noalign{\smallskip}
f$_{1}$	&	9.777	&		&	17.76	&	31	\\
f$_{2}$	&	8.657	&		&	5.95	&	24	\\
f$_{3}$	&	7.015	&		&	3.48	&	19	\\
f$_{4}$	&	8.166	&		&	1.64	&	9.8	\\
f$_{5}$	&	8.880	&		&	1.43	&	9.5	\\
f$_{6}$	&	3.151	&		&	1.24	&	9.5	\\
f$_{7}$	&	1.519	&		&	1.06	&	7.2	\\
f$_{8}$	&	10.495	&		&	1.05	&	8.0	\\
f$_{9}$	&	9.827	&		&	1.04	&	7.7	\\
f$_{10}$	&	2.389	&		&	0.97	&	7.1	\\
f$_{11}$	&	0.763	&	f$_{6}-$f$_{10}$	&	0.88	&	5.5	\\
f$_{12}$	&	12.911	&		&	0.84	&	6.5	\\
f$_{13}$	&	8.522	&		&	0.77	&	5.4	\\
f$_{14}$	&	6.944	&		&	0.78	&	5.5	\\
f$_{15}$	&	4.760	&		&	0.66	&	5.4	\\
f$_{16}$	&	5.722	&		&	0.62	&	4.8	\\
f$_{17}$	&	9.894	&		&	0.65	&	4.8	\\
f$_{18}$	&	1.667	&	f$_{9}-$f$_{4}$	&	0.52	&	4.0	\\
f$_{19}$	&	18.434	&	f$_{1}$+f$_{2}$	&	0.52	&	9.6	\\
f$_{20}$ 	&	16.792	&	f$_{1}$+f$_{3}$	&	0.38	&	6.4	\\
f$_{21}$	&	18.483	&	f$_{2}$+f$_{9}$	&	0.33	&	6.5	\\
f$_{22}$	&	19.554	&	2f$_{1}$	&	0.31	&	6.5	\\
\noalign{\smallskip}
\hline
\end{tabular}
\end{center}
\end{table}

As described in the previous section, the regular frequencies present in the data were analyzed with the {\tt PERIOD04} package. Because of the presence of
both stellar and orbital peaks, we illustrate the intermediate and final results in Fig. \ref{btcnc}, where panels (a) to (f) are discussed below:

(a) The top panel shows the spectral window with the value of amplitude 1 at 0 cd$^{-1}$ off scale to illustrate the aliasing occurring near the orbital frequency of 14.2 cd$^{-1}$ and its multiples.

(b) This shows the Fourier amplitude spectrum from 0 to 50 cd$^{-1}$. Three frequencies marked as ``f1'', ``f2'', and ``f3'', are dominant.

(c) If we inspect the low-amplitude region (with f$_1$ to f$_3$ off-scale), we see a large number of additional frequencies in the 0 - 50 cd$^{-1}$ range.

(d) The number of frequency peaks is severely reduced once we prewhiten f$_1$ to f$_3$. In fact, except for orbital frequencies, which are multiples of 14.19 cd$^{-1}$, no significant peaks remain at frequencies higher than 20 cd$^{-1}$.

(e) However, this still leaves a number of additional significant peaks in the 0 - 20 cd$^{-1}$ range. These peaks are stellar as well as orbital frequencies, artifacts and aliases.

(f) 22 statistically significant frequencies are judged by us to originate in the star. After we prewhiten these 22 frequencies, several low-amplitude peaks remain which we suspect to be mainly nonstellar because they are close to the frequencies of known artifacts. The next stellar peak is found at 7.71 cd$^{-1}$,
but its amplitude signal/noise of 3.7 is below our significance limit. Consequently, the search for stellar frequencies is stopped at this point.

The 22 significant stellar frequencies are listed in Table \ref{freqs}. The average measurement fits the solution to $\pm$ 2.5 millimag. This also holds for the section of the data shown in Fig. \ref{btic}. The listed frequencies should be free of orbital effects with one exception: f$_4$ at 8.166 cd$^{-1}$ is separated by only 1/T from a suspected orbital alias frequency at 8.199 cd$^{-1}$, where T is the length of observations. A misidentification should not be a problem because of the relatively large
amplitude of f$_4$. Furthermore, after prewhitening f$_4$, we are left with a small peak at 8.199 cd$^{-1}$, the expected orbital alias frequency.

How do the new frequencies compare with those found in previous, smaller studies (Breger 1973; Guerrero, Mantegazza \& Scardia 1997; Breger 1980; Kim \& Lee 1995; Freyhammer, Larsen \& Petersen 1997; Hernandez et al. 1998; Pena et al. 1998)?
All studies have detected the dominant frequency of 9.78 cd$^{-1}$, while the 7.02 cd$^{-1}$ frequency usually was also seen.
We interpret the differences in the other reported values for suspected frequencies to be due to the relatively small data sets and 1 cd$^{-1}$ aliasing of earth-based telescope measurements. Examples may be the second and third frequencies reported by Breger (1980), which both are 1 cd$^{-1}$ aliases of frequencies found in the present paper.

\section{EP Cnc (= HD 73819 = KW 348)}

For EP Cnc we have detected 20 stellar frequencies, which are listed in Table \ref{freqsep}.

\begin{table}[htb]
\caption{Pulsation frequencies of EP Cnc for 2008 and 2009}
\label{freqsep}
\begin{center}
\begin{tabular}{lrcrr}
\hline
\noalign{\smallskip}
\multicolumn{3}{c}{Frequency} & Amplitude & S/N\\
& cd$^{-1}$ & & millimag & ratio\\
\noalign{\smallskip}
\hline
\noalign{\smallskip}
& $\pm$ 0.0027$^{1)}$ & & $\pm$ 0.02\\
\noalign{\smallskip}
f$_{1}$	&	5.7741	&		&	1.63	&	14	\\
f$_{2}$	&	6.5198	&		&	1.26	&	13	\\
f$_{3}$	&	5.4610	&		&	1.03	&	10	\\
f$_{4}$	&	1.7712	&		&	1.00	&	7.0	\\
f$_{5}$	&	8.2274	&		&	0.93	&	12	\\
f$_{6}$	&	10.3810	&		&	0.89	&	13	\\
f$_{7}$	&	6.6596	&		&	0.89	&	11	\\
f$_{8}$	&	9.9985		& f$_{4}$+f$_{5}$	&	0.86	&	13	\\
f$_{9}$	&	9.4380	&		&	0.58	&	8.4	\\
f$_{10}$	&	9.7331	&		&	0.49	&	7.6	\\
f$_{11}$	&	5.7276	&		&	0.49	&	5.9	\\
f$_{12}$	&	10.6764	&		&	0.47	&	7.4	\\
f$_{13}$	&	3.8993	&		&	0.42	&	4.2	\\
f$_{14}$	&	6.9675	&		&	0.38	&	5.1	\\
f$_{15}$	&	11.5016$^{2)}$	&	f$_{1}$+f$_{10}$	&	0.38	&	6.2	\\
f$_{16}$	&	6.2907	&		&	0.34	&	4.3	\\
f$_{17}$	&	7.3542	&		&	0.30	&	4.1	\\
f$_{18}$	&	6.5588	&		&	0.30	&	4.0	\\
f$_{19}$	&	7.5968	&		&	0.30	&	4.3	\\
f$_{20}$	&	12.2948	&	f$_{1}$+f$_{2}$	&	0.28	&	4.4	\\
\noalign{\smallskip}
\hline
\noalign{\smallskip}
\end{tabular}
\end{center}
$^{1)}$ The formal errors for the listed frequencies are as small as 0.0001~cd$^{-1}$. However, since annual alias values cannot be excluded, we list the value
of the annual alias.\\
$^{2)}$ The annual alias value of f$_{15}$ at 11.5043 cd$^{-1}$ fits equally well. The alternate value can be identified as a combination frequency (f$_4 $+ f$_{11}$).\\
\end{table}

We also calculated individual amplitudes for 2008 and 2009 separately. The variation of amplitudes between 2008 and 2009 was small with an average deviation of 0.10 millimag, except for the low frequency f$_4$, where the amplitude changed from 0.9 to 1.2 millimag.

The results are in excellent agreement with earlier measurements. The variability of EP Cnc had been discovered by Martin \& Hube (1989)
with an estimate of a 5.46 cd$^{-1}$ frequency. The multisite campaign by the  {\it Delta Scuti Network} determined three frequencies (Breger et al. 1994), which were identical to the three modes with the highest amplitudes in the present data sets.

\section{BS Cnc (HD 73450 = KW 154)}

For BS Cnc we have detected 20 stellar frequencies, which are listed in Table \ref{freqsbs}. Due to the strong amplitude variability associated with some of the detected frequencies, we list the amplitudes from 2008 and 2009 separately. However, the detection significance, S/N, and therefore the decision which frequencies to include, was computed from the combined data set.

An interesting result is the large amplitude variability of two frequencies. The amplitude of 15.53 cd$^{-1}$ varied from 2.35 to 0.86 millimag from 2008 to 2009. The
behavior of the new frequency at 31.60~cd$^{-1}$ is puzzling. The frequency is essentially absent in 2008 and was present with a large amplitude of 2.1 millimag during 2009. Within each of the two observing seasons, the annual amplitude values did not change. Our attempts to identify the frequency as an instrumental artifact was unsuccessful: (i) Since the unreduced raw data from 2009 also showed this frequency, it is not caused by our reduction scheme. (ii) The frequency is 2 cd$^{-1}$ away from an orbital frequency (2f$_\mathrm{orb}$), but the observed amplitude is too high for an alias. (iii) Finally, the frequency value cannot be matched by a combination of another pulsation frequency with the orbital multiples. We conclude that the detection of the stellar 31.60 cd$^{-1}$ frequency is probably real.

We can now compare our frequencies with the known frequencies previously determined from ground-based photometry. The agreement
with the four frequencies determined during the multisite {\tt Stephi} campaign (Hernandez et al. 1998)
 is excellent. In fact, both studies find the same two frequencies
to be dominant. The 34.23 cd$^{-1}$ frequency found by Hernandez et al. (1998)
 (1.4 millimag amplitude) can be matched by 34.41 cd$^{-1}$ (1.4 millimag amplitude) and a smaller-amplitude frequency at 34.20 cd$^{-1}$. Finally, the fourth frequency at 41.63 cd$^{-1}$ agrees with our value of 41.65 cd$^{-1}$, although our amplitude of 0.24 millimag is much smaller.

\begin{table}[htb]
\caption{Pulsation frequencies of BS Cnc for 2008 and 2009}
\label{freqsbs}
\begin{center}
\begin{tabular}{llcrr}
\hline
\noalign{\smallskip}
\multicolumn{2}{c}{Frequency} & \multicolumn{2}{c}{Amplitude} & S/N\\
& & 2008 & 2009 &  ratio\\
\multicolumn{2}{c}{cd$^{-1}$}&  \multicolumn{2}{c}{millimag} \\
\noalign{\smallskip}
\hline
\noalign{\smallskip}
& $\pm$ 0.0027$^{1)}$ & $\pm$ 0.02 & $\pm$ 0.04\\
\noalign{\smallskip}
f$_{1	}$&	17.0363	&	6.23	&	5.72	&	35	\\
f$_{2	}$&	34.4088	&	1.69	&	1.15	&	18	\\
f$_{3	}$&	23.8118	&	1.69	&	1.13	&	23	\\
f$_{4	}$&	15.5272	&	2.35	&	0.86	&	21	\\
f$_{5	}$&	16.9347	&	1.19	&	1.03	&	18	\\
f$_{6	}$&	19.2807	&	0.92	&	0.87	&	17	\\
f$_{7	}$&	19.3379	&	0.88	&	0.76	&	15	\\
f$_{8	}$&	23.3644	&	0.83	&	0.65	&	14	\\
f$_{9	}$&	17.7465	&	0.82	&	0.95	&	14	\\
f$_{10	}$&	31.5988	&	0.20	&	2.10	&	12	\\
f$_{11	}$&	30.2671	&	0.98	&	0.35	&	10	\\
f$_{12	}$&	6.38830	&	0.34	&	0.36	&	8.2	\\
f$_{13	}$&	27.7331	&	0.72	&	0.06	&	8.0	\\
f$_{14	}$&	36.6705	&	0.32	&	0.22	&	7.5	\\
f$_{15	}$&	35.0716	&	0.33	&	0.31	&	7.4	\\
f$_{16	}$&	34.1969	&	0.43	&	0.07	&	7.2	\\
f$_{17	}$&	21.6585	&	0.33	&	0.34	&	7.3	\\
f$_{18	}$&	26.6687	&	0.38	&	0.33	&	6.7	\\
f$_{19	}$&	27.5303	&	0.38	&	0.29	&	6.3	\\
f$_{20	}$&	37.0110	&	0.27	&	0.13	&	6.3	\\
f$_{21	}$&	36.8421	&	0.25	&	0.22	&	6.2	\\
f$_{22	}$&	21.1147$^{2)}$		&	0.26	&	0.32	&	6.1	\\
f$_{23	}$&	32.1866	&	0.40	&	0.29	&	5.9	\\
f$_{24	}$&	7.86910	&	0.27	&	0.17	&	5.8	\\
f$_{25	}$&	21.4462$^{2)}$		&	0.24	&	0.30	&	5.8	\\
f$_{26	}$&	34.8510	&	0.23	&	0.33	&	5.7	\\
f$_{27	}$&	24.4267	&	0.28	&	0.24	&	5.3	\\
f$_{28	}$&	33.9294	&	0.39	&	0.07	&	5.3	\\
f$_{29	}$&	35.1557	&	0.22	&	0.21	&	5.2	\\
f$_{30	}$&	41.6514	&	0.22	&	0.25	&	5.1	\\
f$_{31	}$&	13.9000	&	0.47	&	0.25	&	5.1	\\
f$_{32	}$&	19.9478	&	0.23	&	0.24	&	5.0	\\
f$_{33	}$&	25.7046	&	0.17	&	0.38	&	4.7	\\
f$_{34	}$&	39.2526	&	0.21	&	0.18	&	4.7	\\
\noalign{\smallskip}
\hline
\noalign{\smallskip}
\end{tabular}
\end{center}
$^{1)}$ The formal errors for the listed frequencies are as small as 0.0001~cd$^{-1}$. However, since annual alias values cannot be excluded, we list the value
of the annual alias.\\
$^{2)}$ The sum of these two frequencies is three times the orbital frequency, suggesting possible instrumental effects affecting the two peaks and
their amplitudes. The agreement may be accidental, but some caution is advised.\\
\end{table}

\section{HD 73872 (= KW 375)}

The variability of this cluster star has not been known before. A very brief variability test with photometric observations covering only 2.8 hours showed constancy to 0.002 mag (Breger 1970). However, the new data clearly show variability with small amplitudes. Table \ref{freqshd} lists the 18 frequencies detected by our {\tt MOST} satellite data.

\begin{table}[htb]
\caption{Pulsation frequencies of HD 73872 for 2008}
\label{freqshd}
\begin{center}
\begin{tabular}{lrrr}
\hline
\noalign{\smallskip}
\multicolumn{2}{c}{Frequency} & Amplitude & S/N\\
& cd$^{-1}$ & millimag & ratio\\
\noalign{\smallskip}
\hline
\noalign{\smallskip}
& $\pm$ 0.002 & $\pm$ 0.02\\
\noalign{\smallskip}
f$_{1}$	&	33.416	&	3.03	&	29	\\
f$_{2}$	&	35.981	&	2.53	&	26	\\
f$_{3}$	&	29.807	&	1.52	&	21	\\
f$_{4}$	&	33.542	&	1.50	&	24	\\
f$_{5}$	&	33.786	&	0.88	&	17	\\
f$_{6}$	&	26.086	&	0.81	&	14	\\
f$_{7}$	&	24.657	&	0.76	&   	15	\\
f$_{8}$	&	30.685	&	0.54	&	11	\\
f$_{9}$	&	3.332	&	0.47	&	8.6	\\
f$_{10}$	&	32.215	&	0.46	&	11	\\
f$_{11}$	&	35.824	&	0.46	&	10	\\
f$_{12}$	&	31.016	&	0.45	&	9.5	\\
f$_{13}$	&	18.965	&	0.39	&	9.6	\\
f$_{14}$	&	27.241	&	0.32	&	6.0	\\
f$_{15}$	&	36.337	&	0.27	&	6.1	\\
f$_{16}$	&	31.947	&	0.23	&	5.6	\\
f$_{17}$	&	23.681	&	0.22	&	5.7	\\
f$_{18}$	&	31.714	&	0.19	&	4.6	\\
\noalign{\smallskip}
\hline
\end{tabular}
\end{center}
\end{table}

\section{Discussion}

The four Praesepe stars studied show a wide range of pulsation behavior with frequencies from 0.76 cd$^{-1}$ to 41.7 cd$^{-1}$.
The comparison of observed frequencies with pulsation models (as well as individual stars with each other) requires the determination
of the values of the pulsational constant, $Q$, for each frequency, i.e., the stellar mean
density needs to be known. The $Q$ values are computed from

\setlength{\mathindent}{0pt}
\begin{equation}
\log Q = -\log f + 0.5 \log g + 0.1M_{\mathrm{bol}} + \log T_{\mathrm{eff}} - 6.456,
\end{equation}
where the frequency, $f$, is measured in cd$^{-1}$. For $\delta$~Sct stars, narrowband $uvby\beta$ photometry is
usually used together with model-atmosphere calibrations to derive $g$ and $T_{\mathrm{eff}}$. Due to the uncertainty
of the zero-points of the $uvby\beta$ system and the presence of spectral lines, this procedure
can lead to large uncertainties of up to 25\% in the derived $Q$ values. Fig. 7 of Breger (2000) demonstrates the problem. In this figure, a number of
commonly used calibrations are used to derive a consistency mass, $\cal{M}$, from the atmospheric parameters by

\begin{equation}
{\cal M} \propto  g  L / T^4_{\mathrm{eff}}.
\end{equation}

A comparison of these consistency masses with the expected evolutionary masses shows systematic deviations (up to 50\%) as a function of temperature.
Systematic deviations occur in $all$ available model-atmosphere calibrations which we have tested. Keeping this warning in mind, we shall now calculate the $Q$ values for the frequencies detected in the present study. The absolute magnitudes of the Praesepe stars are accurately known from the revised {\tt HIPPARCOS} parallaxes (Van Leeuwen 2009). We use the unpublished model-atmosphere $uvby\beta$ calibrations by Villa (1998) and the senior author. This calibration
uses $c_1$ zero points to provide an agreement between the observed zero-age-main sequence (ZAMS) with the theoretical $\log g$ values along the ZAMS. This also minimizes the discrepancies between the computed consistency and evolutionary masses. We note that for A stars, our temperatures are close to those of the Moon \& Dworetsky (1985) calibration, while the $\log g$ values are, on the average, higher by 0.15. For the star BS~Cnc, our value of 7600~K is in perfect agreement with the value published by Hernandez et al. (1998).

Our derived values for $T_{\mathrm{eff}}$ and $\log g$ are listed in Table \ref{atmpar}.

\begin{table}[htb]
\caption{Adopted atmospheric parameters for the four Praesepe stars}\label{atmpar}
\begin{center}
\begin{tabular}{lrcrr}
\hline
\noalign{\smallskip}
Star & $T_{\mathrm{eff}}$ in K  & $\log g$ \\
\noalign{\smallskip}
\hline
\noalign{\smallskip}
& $\pm$ 100 & $\pm$ 0.15\\
\noalign{\smallskip}
BT Cnc	&	7370	&	3.45	\\
EP Cnc	&	7950	&	3.70	\\
BS Cnc	&	7600	&	4.20	\\
HD 73872	&	7750	&	4.08	\\
\noalign{\smallskip}
\hline
\end{tabular}
\end{center}
\end{table}

\begin{figure}[htb]
\centering
\includegraphics[width=0.5\textwidth]{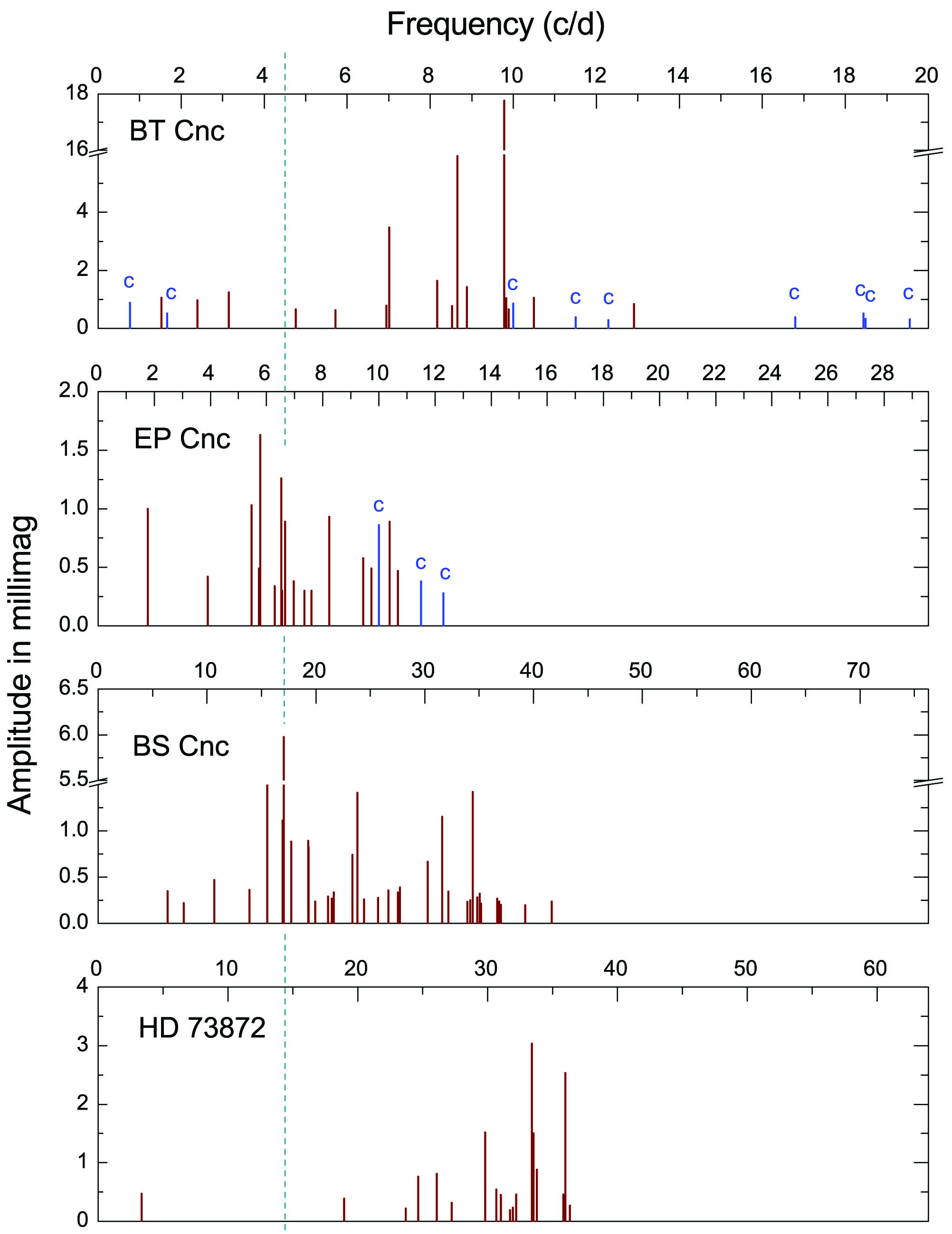}
\caption{Frequency distribution of the four Praesepe stars measured. The frequencies marked with 'c' are combination frequencies, not independent pulsation modes. The dashed vertical line corresponds to the expected frequency of the radial fundamental mode with $Q$ = 0.033d. Furthermore, the frequency scales were chosen in such a way that all expected radial modes are at similar vertical positions.}
\label{fcy}
\end{figure}

We can now normalize the observed frequencies to account for the different mean stellar densities. We use $Q$ = 0.033d, which corresponds to the value for the radial fundamental mode in $\delta$~Sct stars, as our reference value for Fig. \ref{fcy}. Furthermore, the scales on the frequency axes were chosen in such a way that all four panels have the same range in the pulsation constant, $Q$. This figure shows that the four stars exhibit different pulsation behavior, even after these frequency normalizations. Two stars (EP Cnc and BS Cnc) show high-amplitude peaks near the radial fundamental mode, while the other two stars tend to pulsate in higher overtones.

Several low frequencies outside the p-mode range were also detected, e.g., the 1.77 cd$^{-1}$ peak for EP Cnc. Previous earth-based telescope measurements have identified many low frequencies as combination frequencies, f$_i$ $-$ f$_j$, where f$_i$ and f$_j$ are two higher-frequency pressure modes. A beautiful example is the star 44 Tau (Breger \& Lenz 2008). Recent results from space missions such as  {\tt Kepler} have shown that low frequencies can also be a signature of hybrid pulsation, i.e., many $\delta$~Sct stars also pulsate with ($\gamma$-Dor) gravity modes pulsation (Grigahc{\`e}ne at al. 2010). Furthermore, stellar rotation
commonly produces low-frequency peaks, both as f$_\mathrm{rot}$ and 2f$_\mathrm{rot}$ peaks (Balona 2011; Breger 2011). The four Praesepe stars have measured
$v \sin i $ values in a narrow range, from 135 to 156 km\,s$^{-1}$. These values, together with the known absolute magnitudes and temperatures, make it possible to compute lower limits to the rotational frequencies. The lower limits range from 0.64 cd$^{-1}$ for BT Cnc to 1.6 cd$^{-1}$ for HD 73872. An interpretation of the observed low frequencies in terms of stellar rotation can, therefore, not be excluded at this point.

Detailed asteroseismic interpretations are aided by mode identifications of a number of the detected frequencies. Such mode identifications
can be obtained through regular frequency patterns in the frequency spectra, multicolor photometry to give $\ell$ values, 
as well as spectroscopic line-profile analyses. Regular frequency patterns cannot be seen in Fig. \ref{fcy}, the present data are single-color,
and the amplitudes are too small for present-day line-profile analyses. Consequently, we are unable to provide mode identifications at the
present time. The present paper is only a first step of a larger {\tt MOST} study of the Praesepe pulsators. We therefore defer the asteroseismic modeling
to a later paper.

\acknowledgements{

This investigation has been supported
by the Austrian Fonds zur F\"{o}rderung der wissenschaftlichen Forschung through projects P~21830-N16 and P~22691-N16. 
}

\end{document}